\documentclass{PoS}

\usepackage[T1]{fontenc}
\usepackage[latin2]{inputenc}
\usepackage{graphicx}
\input{epsfx}
\usepackage{subfig}
\usepackage{wrapfig}

\usepackage{amsmath}
\usepackage{amsfonts}
\usepackage{amssymb}

\title{Inert Model and evolution of the Universe}

\ShortTitle{Inert Model and evolution of the Universe}

\author{\speaker{Dorota Soko\l owska}%
         \thanks{Work done in collaboration with I.F. Ginzburg, K.A. Kanishev (Novosibirsk University) and M. Krawczyk (University of Warsaw). Work was partly supported by Polish Ministry of Science and Higher Education Grant N N202 230337, FLAVIAnet contract No.MRTN-CT-2006-035482 and EC 6th Framework Programme MRTN-CT-2006-035863.}\\
        University of Warsaw\\
        E-mail: \email{dsok@fuw.edu.pl}}

\abstract{We consider evolution of the Universe after EWSB leading to the present inert phase,
containing a~SM-like Higgs boson and scalar dark particles, among them a Dark Matter
candidate. We address the question, if there is a possibility to
have a sequence of the phase transitions instead of a single one leading directly
from EW symmetric phase to the inert one.}

\FullConference{35th International Conference of High Energy Physics\\
                 July 22-28, 2010\\
                 Paris, France}
\begin{document}

\vspace*{-1cm}
We consider the 2HDM Lagrangian for two $SU(2)$ scalar doublets $\varphi_S,\varphi_D$ with Yukawa interaction set to Model I and the potential $V$:
\[V = -\frac{m_{11}^2}{2} |\varphi_S|^2 -\frac{m_{22}^2}{2}  |\varphi_D|^2+ \frac{\lambda_1}{2}|\varphi_S|^4 +\frac{\lambda_2}{2}|\varphi_D|^4+ \lambda_3|\varphi_S|^2|\varphi_D|^2 +\lambda_4|\varphi_S^\dagger \varphi_D|^2 +\frac{1}{2}\left(\lambda_5 (\varphi_S^\dagger \varphi_D)^2+h.c\right).\]
%
$V$ is invariant under a $Z_2$ transformation: $\varphi_S \rightarrow \varphi_S,\, \varphi_D \rightarrow -\varphi_D$. In the Inert Model also vacuum state is $Z_2$-symmetric  
and the Dark Matter candidate, from the $Z_2$-odd doublet $\varphi_D$, appears.

The most general EWSB solution
$\langle\varphi_S\rangle^T = \frac{1}{\sqrt{2}} (0, v_S), \langle\varphi_D\rangle^T =\frac{1}{\sqrt{2}} (u, v_D)$ gives three neutral extrema ($u=0$):
\textbf{\textit{inert}} $(I_1; v_D=0, v_S^2=v^2)$ with SM-like Higgs $h$ from $\varphi_S$ and DM candidate  from $\varphi_D$ (eg. $H$); \textbf{\textit{inert-like}} $ (I_2; v_S=0, v_D^2=v^2)$ with massless fermions (Model I: only $\varphi_S$ couples to fermions) and no candidate for DM; \textbf{\textit{mixed}} $(M; v_D, v_S \not=0,v^2=v_S^2+v_D^2)$ -- a standard 2HDM extremum. The lowest energy extremum, fulfilling positivity constraints, is the vacuum.

The one-loop thermal corrections to $V$ are $m_{ii}^2(T)=m_{ii}^2-c_iT^2\,\,(i=1,2)\,$ where $c_i=c_i (\lambda_{1-4}; g,g';\\g_t^2+g_b^2 \textrm{ for } i=1)$, with fixed $\lambda_i$ ($g, g^{\prime}$ -- EW gauge couplings, $g_t,g_b$ -- SM Yukawa couplings).

\begin{figure}[hb]
\vspace{-10pt}
  \centering
  \subfloat[$\lambda_{345}>\sqrt{\lambda_1 \lambda_2}$]{\label{fig:wykresa}\includegraphics[width=0.2\textwidth]{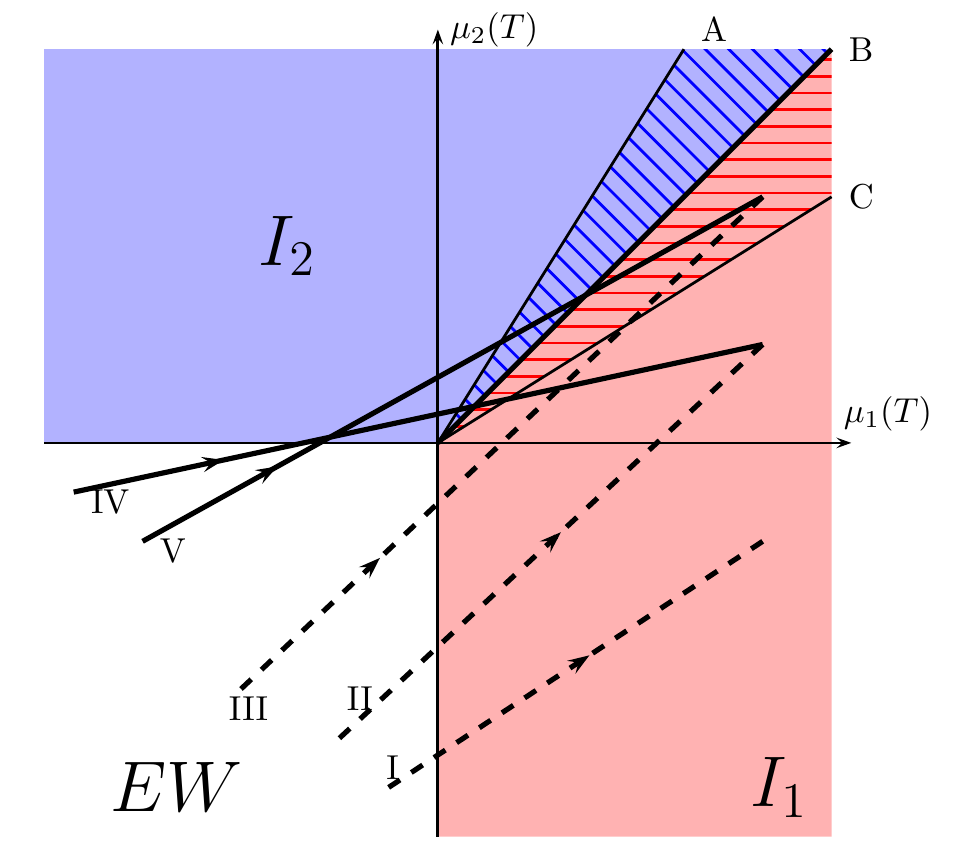}} \qquad
  \subfloat[$0<\lambda_{345}<\sqrt{\lambda_1 \lambda_2}$]{\label{fig:wykresb}\includegraphics[width=0.2\textwidth]{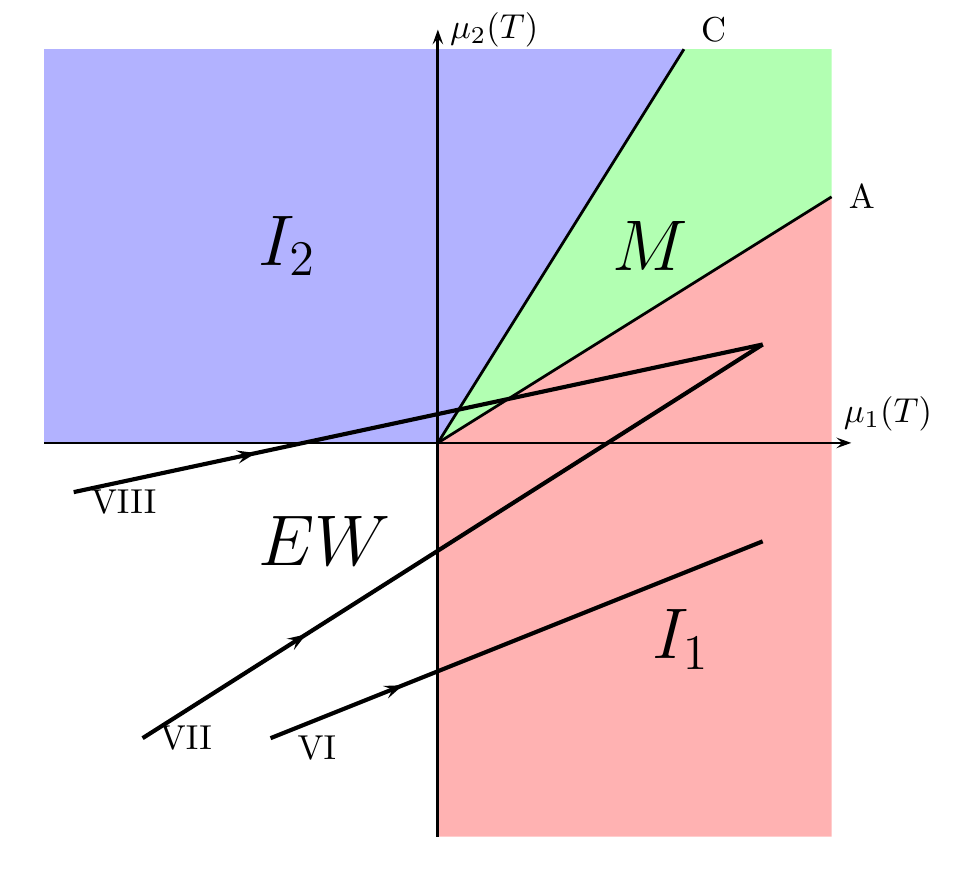}} \qquad
  \subfloat[$-\sqrt{\lambda_1 \lambda_2}<\lambda_{345}<0$]{\label{fig:wykresc}\includegraphics[width=0.2\textwidth]{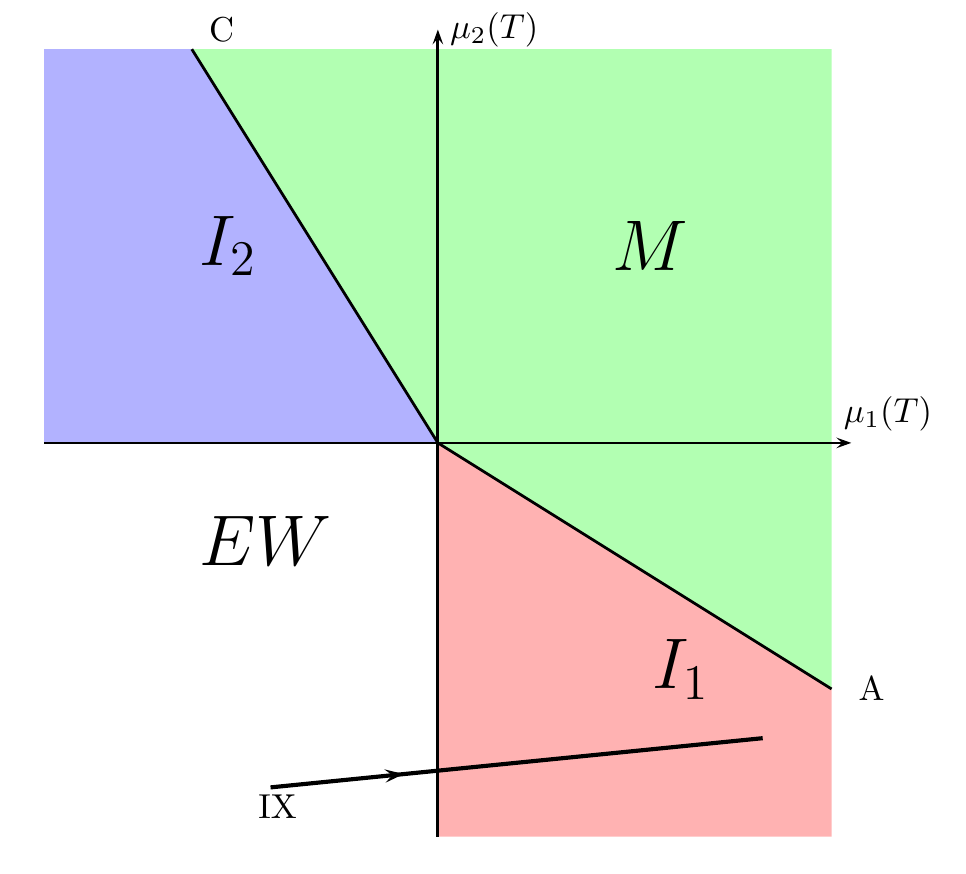}}
\vspace{-5pt}
  \caption{Possible rays from EW symmetric phase to the inert phase $I_1$; $\lambda_{345} = \lambda_{3} + \lambda_{4} + \lambda_{5}$.}
\vspace{-10pt}
  \subfloat[with fermions]{\label{fig:wykresd}\includegraphics[width=0.2\textwidth]{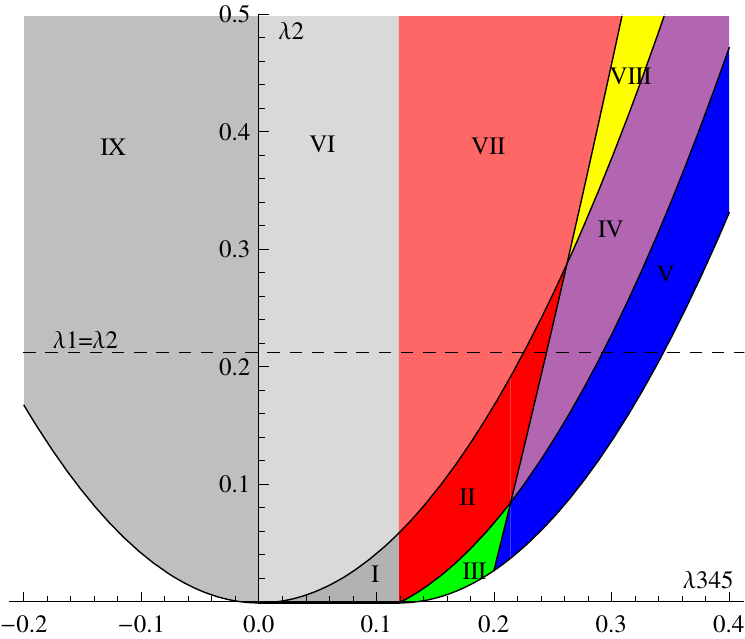}}\qquad
  \subfloat[without fermions]{\label{fig:wykrese}\includegraphics[width=0.2\textwidth]{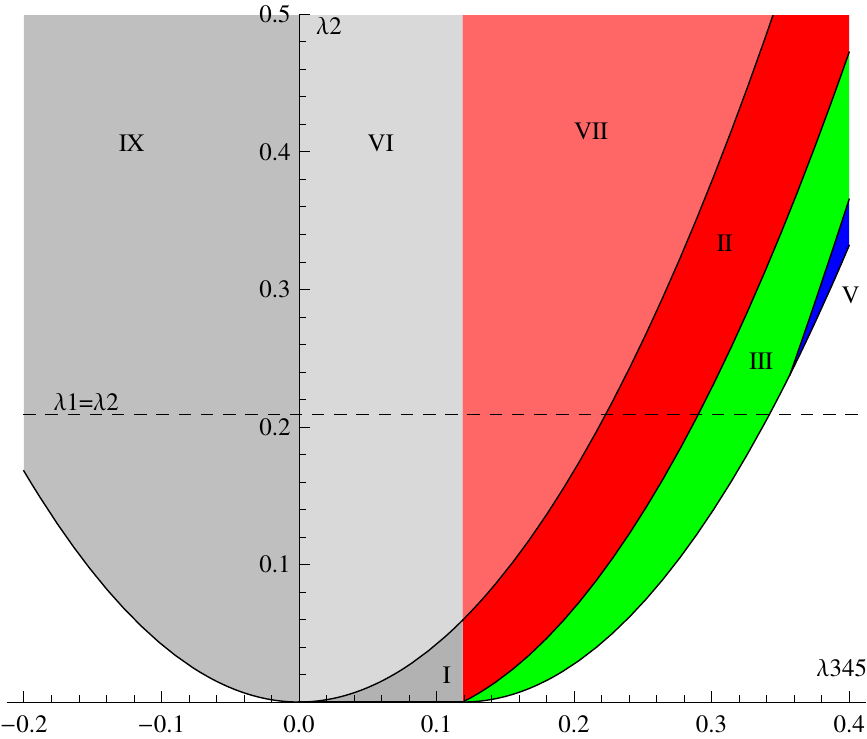}}
\vspace{-5pt}
  \caption{Example of rays for $M_h = 120 \textrm{ GeV}, M_H = 60 \textrm{ GeV}, M_A = 68 \textrm{ GeV}, M_{H^\pm} = 110 \textrm{ GeV}$.}
  \label{fig:rays}
\end{figure}

The possible sequences of phase transitions between different vacua are shown as rays in plots (\ref{fig:wykresa}-\ref{fig:wykresc}), for $(\mu_1, \mu_2)$ plane ($\mu_i(T)=m_{ii}^2(T)/\sqrt{\lambda_i} , i=1,2$). For $EW \to I_1$, $I_1$ is the only vacuum that existed after EWSB. For \textbf{rays I, VI, IX} $I_2$ is not an extremum, for \textbf{rays II, VII} $I_2$ is an extremum, but not a local minimum; for \textbf{ray III} $I_2$ is a local minimum. 
The sequence: $EW \to I_2 \to I_1$, with transition between $I_2$ and $I_1$ vacuum, is possible for \textbf{ray IV}, where $I_2$ is not a local minimum and for \textbf{ray V}, where $I_2$ is a local minimum, that coexists with the global minimum $I_1$. For \textbf{ray VIII} there is a possibility of going through the mixed vacuum $EW \to I_2 \to M \to I_1$. In this case there is only one minimum at any temperature.


Plots \ref{fig:wykresd},\ref{fig:wykrese} show the different vacua sequences (rays I-IX) leading to $I_1$ in ($\lambda_2$,$\lambda_{345}$) plane. For $\lambda_2 \le \lambda_1$ different types of vacua in the past are possible only if the fermionic part of $c_1$ is included.


\begin{thebibliography}{99}
\bibitem{1} \textit{Evolution of Universe to the present inert phase}, I.F. Ginzburg, K.A. Kanishev, M. Krawczyk, D.~Soko\l owska [arXiv:1009.4593]
\end{thebibliography}
\end{document}